\documentclass[prb,aps,groupedaddress,amsmath,twocolumn,showpacs]{revtex4}

\usepackage[dvips]{graphicx}
\usepackage{dcolumn}
\usepackage{bm}
\usepackage[version=3]{mhchem} 
\usepackage{color}

\begin{document}




\title{Finite size scaling with modified boundary conditions} 
\author{Sandro Sorella}
\email[]{sorella@sissa.it}
\affiliation{SISSA -- International School for Advanced Studies, Via Bonomea 265
  34136 Trieste, Italy} \affiliation{Democritos Simulation Center
  CNR--IOM Istituto Officina dei Materiali, 34151 Trieste, Italy} 
\affiliation{Computational Materials Science Research Team, 
RIKEN Advanced Institute for Computational Science (AICS), 
Kobe, Hyogo 650-0047, Japan}
\date{\today}

\begin{abstract}
An efficient  scheme is introduced for a fast and smooth convergence to the thermodynamic limit with finite size cluster calculations.
This is obtained by modifying the energy levels of the non interacting Hamiltonian in a way consistent with the  corresponding  one particle density of states in the thermodynamic limit.  
After this modification exact 
free electron  energies  are  obtained with finite size calculations and  
 for particular fillings that satisfy the so called "closed shell condition".
In this case the ''sign problem'' is particularly mild in the auxiliary field quantum Monte Carlo technique and therefore, with this technique, it is possible to  
 obtain converged energies for the Hubbard model even for $U>0$.
We provide a strong numerical evidence  that phase separation occurs in the low doping region and moderate $U\lesssim 4t$ regime of this model. 
\end{abstract}
\maketitle
After several years of scientific 
effort, based on advanced analytical and numerical methods, only 
very few properties of the 2D Hubbard model have been settled.
The 2D Hubbard model is defined in a 
square lattice  containing a finite number $L$ ($N_h$) of sites (holes):
\begin{equation}
H =K+V= -t \sum_{<i,j>,\sigma} c^{\dag}_{i,\sigma} c_{j,\sigma} + U \sum_i n_{i}^\uparrow n_i^\downarrow
\end{equation}
with standard notations, the kinetic energy operator 
$K$ being equal to $H$  for $U=0$. 
In the thermodynamic limit, namely for   $L\to \infty$ 
at given doping $\delta=N_h/L$ fundamental issues  
such as the existence of a ferromagnetic phase at large 
$U/t$ ratio and/or the stability of an  homogeneous ground state
with possible d-wave superconducting  properties  are 
still  highly debated,  as several approximate numerical techniques  lead 
to controversial and often conflicting  results. 
This situation is particularly important right now, since  
recent progress in the realization of fermionic optical lattices 
could lead  
to the experimental realization of the fermionic Hubbard model.

{\bf Method:} 
We consider the square lattice and generic finite clusters
 that satisfy all the symmetries of the infinite system.
As well known finite square lattices  
  can be defined by two integers $n,m$ such 
that $n^2+m^2=L$, obtained by supercell translation vectors 
$\tau_x =(n,m)$ and $\tau_y=(-m,n)$.
In order to fulfill rotation symmetries, two sequences can be defined:
\begin{eqnarray}
m=0 & L=n\times n & {\rm The ~ usual~sequence} \label{sequence} \\
m=n & L=2 n^2 & {\rm The ~45^0~ degrees~ tilted~ sequence} \nonumber
\end{eqnarray}
On the other hand translation symmetries are recovered by employing periodic
or antiperiodic boundary conditions, the same 
 in both directions $\tau_x$ and  $\tau_y$ 
in order to preserve rotation and reflection symmetries.

In the following we would like to consider the most useful sequence of clusters for converging as fast as possible to the thermodynamic limit.
A simple technique, well known in strongly correlated lattice models\cite{tabc_gros} 
 and in realistic calculations\cite{tabc} 
 is to consider the twisted averaged boundary conditions (TABC) method.
This technique allows an  exact evaluation of converged 
thermodynamic quantities (energy, density matrix, etc.) in the 
non interacting $U=0$ limit.\cite{tabc_gros}  Indeed it has been proven very successful to reduce substantially the finite size effects in  several correlated systems.

In the following we follow a different approach  analogous to TABC in the 
requirement to remove finite size effects in the non interacting limit. 
However the proposed approach  can 
be used more efficiently 
 in combination with the auxiliary field quantum Monte Carlo (AFQMC) method.\cite{hirsch,hst} 
The latter technique is one 
 of the most powerful ones used so far for the study of the Hubbard model, 
as it can project out from a mean-field (Slater determinant) 
state $|MF\rangle$, the exact ground state of the Hamiltonian by the application 
of the imaginary time propagation $\exp(-\tau H)$, for large $\tau$.
Here we consider variational expectation values (Var) on the $\exp(-\tau/2 H) | MF \rangle$ state and non variational mixed estimators (no Var) 
 between the previous state and $\exp(-\tau/2 H) |\psi_0\rangle$, $|\psi_0\rangle$ 
being the ground state Slater determinant for $U=0$. Both quantities clearly converge to exact values for large $\tau$\cite{honey}.  

The sign problem occurs at finite doping and $U/t>0$ but it is particularly mild 
when i) the $U=0$ Hamiltonian has a non degenerate ground state, a situation 
that occurs for particular fillings- the closed shell fillings-  in any finite clusters, ii) the auxiliary 
field transformation is real, as with the proposed approach it is not necessary 
to sample a complex phase. 
Until now several reliable and ''numerically exact'' calculations have been 
performed by means of this technique 
on moderately large clusters (i.e. up to $\simeq 100$ sites and $U/t\lesssim 4t$)\cite{lhst,imada,zhang,jarrell} but it was  difficult to establish  thermodynamically converged results, especially in the weakly correlated regime. 
One should also mention that, recently, a remarkable progress has been made, allowing the complete removal of the time discretization error in the propagator\cite{ct,lhst,ct_recent}, a development that will not be used here, as we have preferred to use a small enough time step $\Delta \tau$, such that this systematic error is negligible, at least as far as the ground state energy is concerned.

Let us now introduce the method.
Consider a finite cluster. The kinetic energy can be written in Fourier space, 
by collecting $k$ points related by point symmetries to the same energy shell
$\epsilon_i$:
\begin{equation} \label{defkin}
K=\sum_{i,\sigma} \epsilon_i \sum_{k_i| \epsilon_k=\epsilon_i} c^{\dag}_{k,\sigma} c_{k,\sigma}
\end{equation}
where $\epsilon_k=-2 t (\cos k_x + \cos k_y)$ in 2D.
We can assume that the energy levels $\epsilon_i$ are defined in ascending order.
Each shell occurs with some multiplicity $g_i$ and:
\begin{equation}
L=\sum\limits_{i=1}^p g_i
\end{equation} 
where $p$ is the number of different energy levels.
In the square lattice case we can choose for simplicity clusters 
that do not contain accidental degeneracies ($g_i \le 8$), namely they 
 are given by 
the second sequence given in Eq.(\ref{sequence}) with PBC (APBC) 
for $n$ odd (even). 
The fundamental quantity that we are 
going to use in order to achieve more easily the thermodynamic limit 
is  the density of states $N(E)$:  
\begin{equation}
N(E) = \int {dk^d \over (2 \pi)^d} \delta (E-\epsilon_k)
\end{equation}
defined in a way that $\int N(E)dE=1$.
This function can be evaluated analytically and/or  computed with arbitrary 
accuracy in the thermodynamic limit for any given lattice model.
We partition the energy bandwidth of the lattice (e.g. $-4t < E < 4t$ in 
the square lattice) in intervals $\bar \epsilon_i$ such that:
\begin{equation} \label{defebar}
g_i/L = \int\limits_{\bar \epsilon_{i-1}}^{\bar \epsilon_{i}} N(E) dE
\end{equation}
The above equation define all 
the levels $\bar \epsilon_i$ by simple induction, because once we know 
, e.g. $\bar \epsilon_{n}$  we can solve the above equation for $i=n+1$ 
and we can obtain univocally $\bar \epsilon_{n+1}$. 
Therefore by setting $\bar \epsilon_0$ equal to the lowest one electron energy 
($-4t$ in the 2D square lattice), all levels $\bar \epsilon_i$ can be 
computed and their level spacing is in exact correspondence with the density 
of states.  
Notice that for the particular symmetry of the DOS in bipartite lattices 
$N(E)=N(-E)$, due to particle-hole
symmetry, it follows that one of the levels
is exactly vanishing.

After the above decomposition, 
in order to fulfill the requirement to have an exact energy for $U=0$ we 
can modify the energy levels of the kinetic energy 
$\epsilon_i \to \tilde \epsilon_i$ in the following way:
\begin{equation} \label{newlevels}
{\bf \tilde \epsilon}_i = { \int \limits_{\bar \epsilon_{i-1}}^{\bar \epsilon_i} 
 E  N(E) dE \over \int \limits_{\bar \epsilon_{i-1}}^{\bar \epsilon_i} 
  N(E) dE }   =  L/g_i  \int \limits_{\bar \epsilon_{i-1}}^{\bar \epsilon_i} E  N(E) dE
\end{equation}
where the latter equality comes just from the definition in Eq.(\ref{defebar}).
In this way the revised  kinetic energy $K \to \bar K$  is 
obtained by replacing $\epsilon_i$ with $\tilde \epsilon_i$ in Eq.(\ref{defkin}).
It is immediate to show that,  when we satisfy 
the closed shell condition, i.e. $N= \sum_{i\le i_F} g_i$, within  these modified 
boundary conditions (MBC) we obtain straightforwardly that the ground state 
energy per site is:
\begin{equation}
 <K/L> = 2  \sum\limits_{i \le i_F} {g_i \over L}  {\bf \tilde \epsilon}_i = 2
 \int\limits_{\bar \epsilon_0}^{\bar \epsilon_{i_F} } E N(E) dE
\end{equation}
namely the exact energy per site for $U=0$ at the thermodynamic density:
$$N/L=2 \int \limits_{\bar \epsilon_0}^{\bar \epsilon_{i_F}} N(E) dE$$
, where the factor two in the above equation takes into 
account  the spin components. 
 
At finite $U$ it is quite simple to show that the above sequence of 
lattices with MBC converges to the exact  thermodynamic limit 
 because for large $L$ the modification of the levels, as compared to the 
original ones, becomes irrelevant.

In this way we have several advantages and simplifications:
\begin{itemize}
\item The Hamiltonian   is always real with MBC (a positive 
property for the sign problem).\cite{zhang_new}
\item The non interacting exact limit is obtained for the closed
shell fillings. Thus we expect less size effects just for those particular densities  less affected by the  sign problem within AFQMC.
\item The MBC  satisfy  all the symmetries of the infinite systems.
For instance when we apply TABC, each boundary with 
a non zero twist generally breaks all point spatial group symmetries, maintaining only 
translation symmetry. This may not affect the average result, but it becomes certainly more difficult to converge to the exact result, i.e. one needs larger projection times in AFQMC 
due to smaller finite size gaps that occur after a small symmetry breaking 
perturbation of the Hamiltonian given by a tiny twist of the boundary conditions.
\item Last but not least, MBC are  rather trivial to implement in the AFQMC as it is enough to change the propagator 
$\exp(-\Delta \tau K )\to \exp(-\Delta \tau \bar K )$.
This matrix is never sparse and 
has to be  computed in advance within AFQMC for its efficient 
implementation. Thus the use of MBC does not lead to  any overhead in the performances  of the algorithm.
\end{itemize}
\begin{figure}[h]
\begin{center}
\includegraphics[scale=0.5]{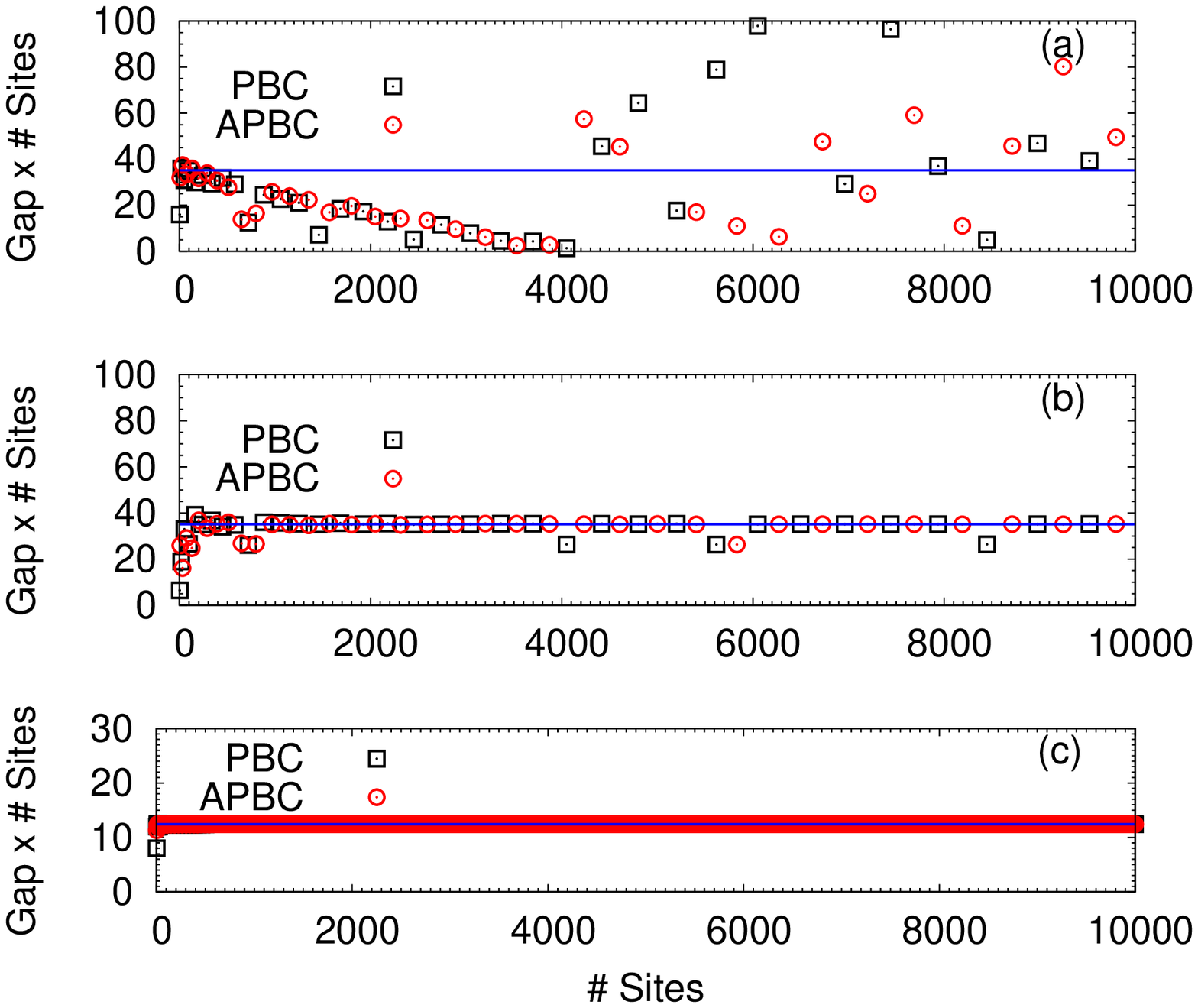}
\includegraphics[scale=0.45]{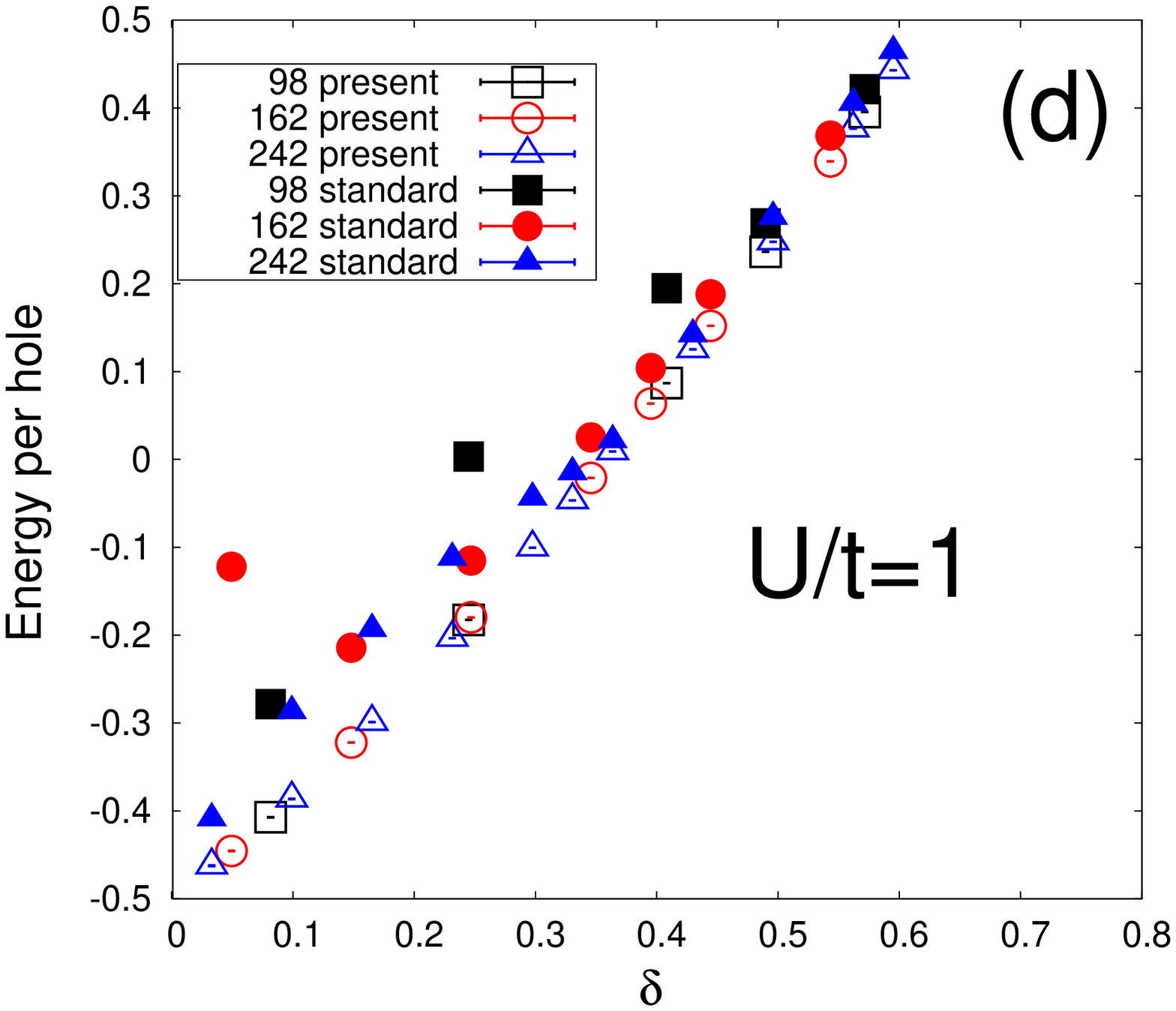}
\caption{ $U=0$ energy gap  for finite clusters  
with non degenerate ground state at the closest filling $N/L=0.9$ with
$45^0$ tilted  PBC or APBC. 
This gap  is multiplied by the number sites, as it should converge 
for $L\to \infty$  to $g/N(\mu)$,
where $\mu$  is the thermodynamic chemical potential at this filling and $g$ is the multiplicity 
of the energy level ($g=8$ in 2D and $g=2$ in 1D).
  (a): standard 2D clusters. (b) modified 2D clusters according to this work (see text). (c): standard 1d clusters. (d) Energy per hole for standard clusters with PBC (filled symbols) and with  MBC  (empty symbols).
}
\label{fig:gap}
\end{center}
\end{figure}

\begin{figure}[h]
\begin{center}
\includegraphics[scale=0.5]{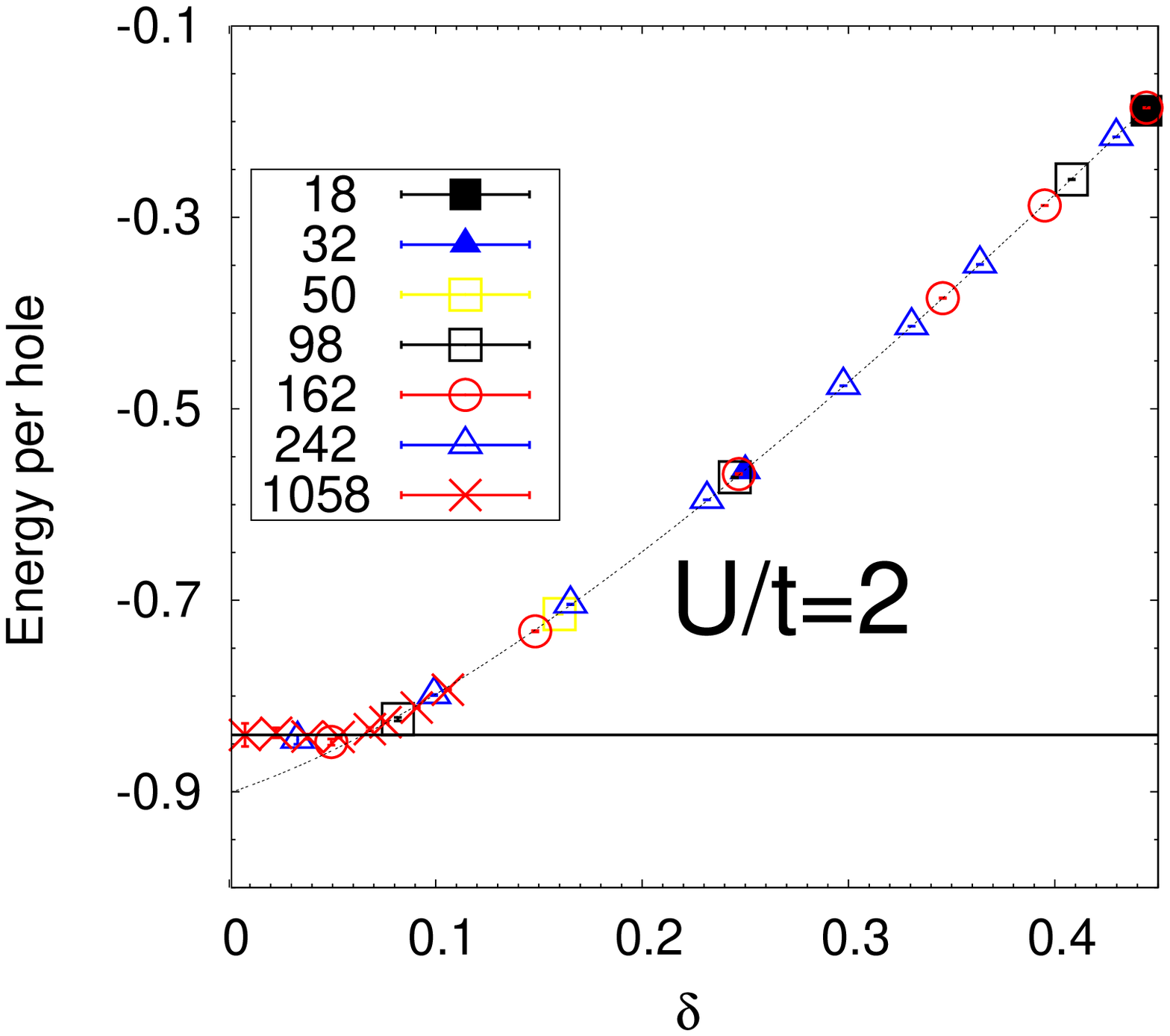}
\includegraphics[scale=0.5]{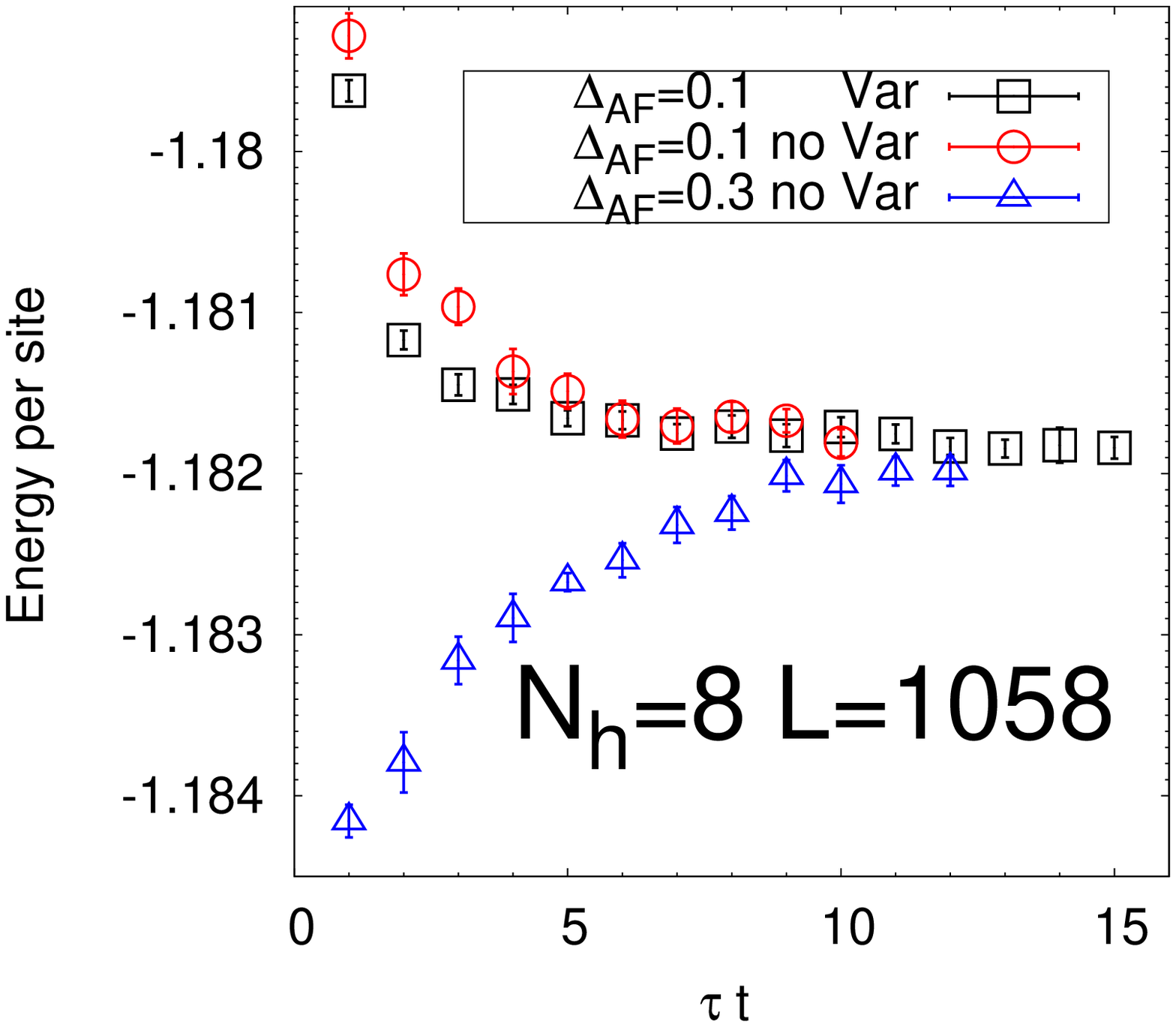}
\caption{ Upper panel: Energy per hole in the 2D Hubbard model. Finite size effects are rather well behaved by using MBC and at small doping the energy per hole approaches almost exactly an horizontal line. This implies phase separation as discussed in the text.  The dashed line is a fit (see table) of the data for $\delta > 6.5 \%$.
 Lower panel: energy vs $\tau$ convergence starting with several 
initial left and right wave functions. "Var" ("no Var") stands for the (non-)variational calculation (see text) with different values of the 
 antiferromagnetic  parameter $\Delta_{AF}$ in the mean-field determinant.  
}
\label{fig:u2}
\end{center}
\end{figure}
As it is shown in Fig.(\ref{fig:gap}a-c), the most important problem in the usual 
sequence of finite clusters is that the $U=0$ finite size gap behaves erratically when $L$ increases, in spatial dimensionalities $D>1$ and away from 
commensurate fillings. This precludes to obtain accurate extrapolations to the 
thermodynamic limit at least for moderate $U/t$ and finite dopings (see e.g. Fig.\ref{fig:gap}d). As it is evident in Fig.(\ref{fig:gap}a-b) 
 the MBC behave much better in this respect.
Only few clusters  scatter from the converged $\simeq g/N(E)$ energy level separations, but they correspond to atypical lattices when $g_i \ne 8$ at the highest occupied or lowest unoccupied free-electron energy levels. Remarkably a reasonably converged value of this quantity occurs only after few hundreds sites, which is feasible for a numerical approach (see 
e.g. Fig.~\ref{fig:gap}d).

{\bf Results:} we have carried out a systematic finite size scaling study of the energy per hole
in the moderate $U/t$ regime. As pointed out in the milestone paper by Emery and Kivelson\cite{emery} a minimum at doping  $\delta_c$ in the energy per hole, corresponding to a given variational ansatz, implies its instability against phase separation because it is possible to gain energy for $\delta \le \delta_c$  by segregating the holes in a hole rich region with the same type of ansatz.
In an exact calculation, whenever it is possible to carry out the thermodynamic limit, clearly we have to obtain a constant energy per hole in all the region 
$\delta \le \delta_c$, just because the compressibility - e.g.  
the slope in the energy per hole
at $\delta \to 0$-, cannot be negative in an exact calculation. 
As shown in Fig.~\ref{fig:u2}(a), at $U/t=2$ we can safely reach the thermodynamic limit with the clusters considered, thanks to the very small finite size effects 
introduced by the method proposed in the previous section, that, in this case, 
 look considerably smaller than standard TABC\cite{zhang}. The convergence 
in imaginary time is quite clear also for the most difficult case at small doping (see Fig.\ref{fig:u2}b) and also independent of the initial trial function used. 
In this case we have used a mean field state with a non zero antiferromagnetic 
order parameter $\Delta_{AF}$ along the $x-$ spin direction.\cite{lhst} 
 In the more difficult cases for larger $U/t$ we have also used a d-wave order 
parameter $\Delta^{BCS}_{x^2-y^2}$ in order to converge faster or at least 
for  obtaining the lowest possible variational energies compatible with a reasonable average sign $<s>\ge 0.05$. It is clear that this by no means implies the 
existence of a non zero superconducting order parameter in the ground state, an issue that will not be discussed in the present work.     

At $U/t=2$  the achieved flat behavior of the 
energy per hole for $\delta \le 6.5\%$ clearly indicates the accuracy of the AFQMC at this small coupling, that  is indeed able to determine  phase separation  just by imaginary time projection of an homogeneous trial state. 
At larger coupling (see Ref.\onlinecite{epaps}), though we have not been able to reach the same cluster size 
and the same length of the projection times, the accuracy in the energy per hole
appears acceptable and allow us to determine the phase separated region 
for $U\le 4 t$ (see table).  
Notice that, in this table, the energy 
gain to have phase separation can be measured 
by the difference of the minimum hole energy $E^h_{min}$ obtained for the 
largest clusters at doping $\delta \le \delta_c$ and the hole energy 
$a_0$ extrapolated at $\delta=0$ using only doping values  clearly outside the 
phase separated region.
In other words $a_0$ represents the energy per hole  of the uniform phase extrapolated at zero doping.
This difference $a_0-E^h_{min}$ appears to be very large $\simeq 0.1t$ at $U/t=2,3$ and less evident for $U/t=4$. This is probably the reason why at $U/t=4$, there have been several controversial claims\cite{lhst,jarrell,becca,zhang}.
 Given this behavior, it is also possible that, at larger $U/t$, the phase 
separation may be less evident and $\delta_c$ may 
significantly decrease\cite{tocchio}, despite some works indicate exactly the opposite effect\cite{imadaprb,zhang}. It is clear that 
 at large $U/t$ this important issue remains still open.
 On the other hand, at $U/t=1$, we have not obtained evidence of phase separation, because probably we cannot reach enough small doping values with the affordable finite clusters $L\lesssim 1058$.  Indeed at this coupling value also the antiferromagnetic order parameter $m_{AF}$  cannot be detected numerically, being extremely small even at $\delta=0$. Since the existence of antiferromagnetism is at the basis of the phase separation argument\cite{emery} it is possible 
that also $\delta_c$ can be exponentially small at small $U/t$,  as is the case for 
$m_{AF} \simeq \exp (-\propto 1/\sqrt{t/U}) $ within the Hartree-Fock theory\cite{hirsch}. 

\begin{widetext}
\begin{table}[h]
\begin{tabular}{|c|l|l|l|l|l|l|l|l|l|}
\hline
 U/t & $E^h_{min}$ & $\delta_{c}$ & $a_0$ & $a_1$ & $a_2$ & $a_3$ & $a_4$ & $a_5$ & $\Delta_{max}$  \\
\hline
 1 &  -0.484(13)  &  0.00(1) & -0.49185 & 0.88825 & 1.99156 & -2.46827 & 2.20290 & -0.73928 & 0.00053 \\
\hline
 2 &  -0.843(1) &  0.067(5) & -0.90164 & 0.73469 & 3.45846 & -4.87765 & 4.03989  & -1.27852 & 0.00062  \\
\hline
 3 &  -1.125(1) & 0.105(10)&  1.20909 & 0.41515  & 4.84506 & -6.43761 & 4.74676 & -1.35904 & 
0.00034 \\ 
\hline
 4 &  -1.342(2) & 0.110(15) & -1.35005 & -0.78626 & 9.18032 & -13.03818 & 9.60894 & -2.75519 & 0.0012\\
\hline
\end{tabular}
\caption{ Estimated energy per hole in the thermodynamic limit.
 The functional form of the fit is a $5^{th}$ order polynomial $E^h(\delta) =\sum\limits_{i=0}^5 a_i \delta^i$ determined by the largest cluster data in the region $\delta_{c} < \delta \le 1$. The rightmost column represents the maximum error of the fit for $\delta \times E_h(\delta)$ (the energy per site referenced to the undoped case). $E^h_{min}$ represents the estimated minimum energy per hole for $\delta \to 0$. Number(s) between  brackets indicate error bars in the last digit(s).}
\end{table}
\end{widetext}
{\bf Conclusions:}
We have introduced a technique for controlling finite size effects in an efficient way, an approach particularly suited for the AFQMC method.
In this way a strong numerical evidence is given  that phase separation is robust at small dopings  and  $U/t$ values. With this approach it is  possible to study other possible 
phases\cite{prokovev_last,metzner,castellani}, with a better control of finite size effects at incommensurate dopings and weak 
couplings, as well as it is possible to export the method to other techniques, 
that may have problem of convergence to the thermodynamic limit especially at weak couplings.
Indeed we have preliminary verified that, by choosing boundary conditions that break the 
symmetry of the lattice (e.g. cylindrical),  much better results (i.e. much closer to the thermodynamic limit) can be obtained 
by correcting the energy levels of the $U=0$ Hamiltonian according to the proposed method.
Finally we want to remark that  the method can be easily extended to realistic calculations that do not explicitly require a local Hamiltonian\cite{alavi,cpqmc} as in AFQMC. 
This can be achieved  
by considering as an input for the correlated calculation the band-resolved 
DOS obtained with an uncorrelated Hamiltonian, such as the Khon-Sham one in Density Functional Theory. The same technique as above can be used to reduce finite size effects by preserving charge neutrality even in presence of the Coulomb long range interaction, a property that is difficult to fulfill with TABC, if we require that the non interacting limit should remain exact with a finite supercell calculation. 

{\bf Acknowledgments} I  acknowledge useful discussions with F. Becca, S. Yunoki, and L. Tocchio, and support by AICS Riken  Kobe and by MIUR COFIN 2010.
Part of this research has used computational resources of the 
K computer provided by the RIKEN Advanced Institute for Computational 
Science through the HPCI System Research Projects 
(hp120174 and hp140092) and  CINECA ISCRA (grant:HP10C0DZUP).




\end{document}


\widetext
This supplementary material contains energies of the Hubbard model 
with the AFQMC method described in the paper for various cluster sizes 
with periodic boundary conditions rotated by $45$ degrees.
These square lattices  have an even number of sites $L= 2 n^2$ where $n$ is an 
odd (even) integer with PBC (APBC), and fulfill all rotation symmetries of the infinite lattice.
In all the tables error bars are between brackets with usual conventions.
In all the forthcoming results 
the guiding function is defined by means of the ground state 
$\psi_{MF}$ of the following mean field Hamiltonian:
\begin{equation}
H_{MF} = \bar K - \mu_0 N +\left[ 
\Delta_{AF} \sum_R (-1)^{x+y} c^{\dag}_{R,\uparrow} c_{R,\downarrow} +\Delta_{BCS}^{x^2-y^2} \sum_k (\cos k_x - \cos k_y ) c^{\dag}_{k,\uparrow} c^{\dag}_{-k,\downarrow} + {\rm h.c.} \right]
\end{equation}
where $\bar K$ is the modified kinetic energy with MBC  
and $\mu_0\ne 0$ is used when $\Delta_{BCS}^{x^2-y^2} \ne 0$ as the non interacting 
chemical potential value, namely at the middle of the non interacting highest energy 
 occupied level $\bar \epsilon_{i_F+1}$ and lowest energy unoccupied level 
$\bar \epsilon_{i_F}$, $\mu_0={ \bar \epsilon_{i_F+1} + \bar \epsilon_{i_F})\over 2}$, $N=\sum\limits_{k,\sigma}  c^{\dag}_{k,\sigma} c_{k,\sigma}$ being the 
total number of particle operator that is not modified within the MBC approach. 
 $R=(x,y)$ is a lattice point belonging to this  lattice, namely 
 $R=(x,y)$ is equivalent to $(x\pm n,y+n)$. 

\begin{figure}[ht]
\centering
\includegraphics[width=1.0\columnwidth]{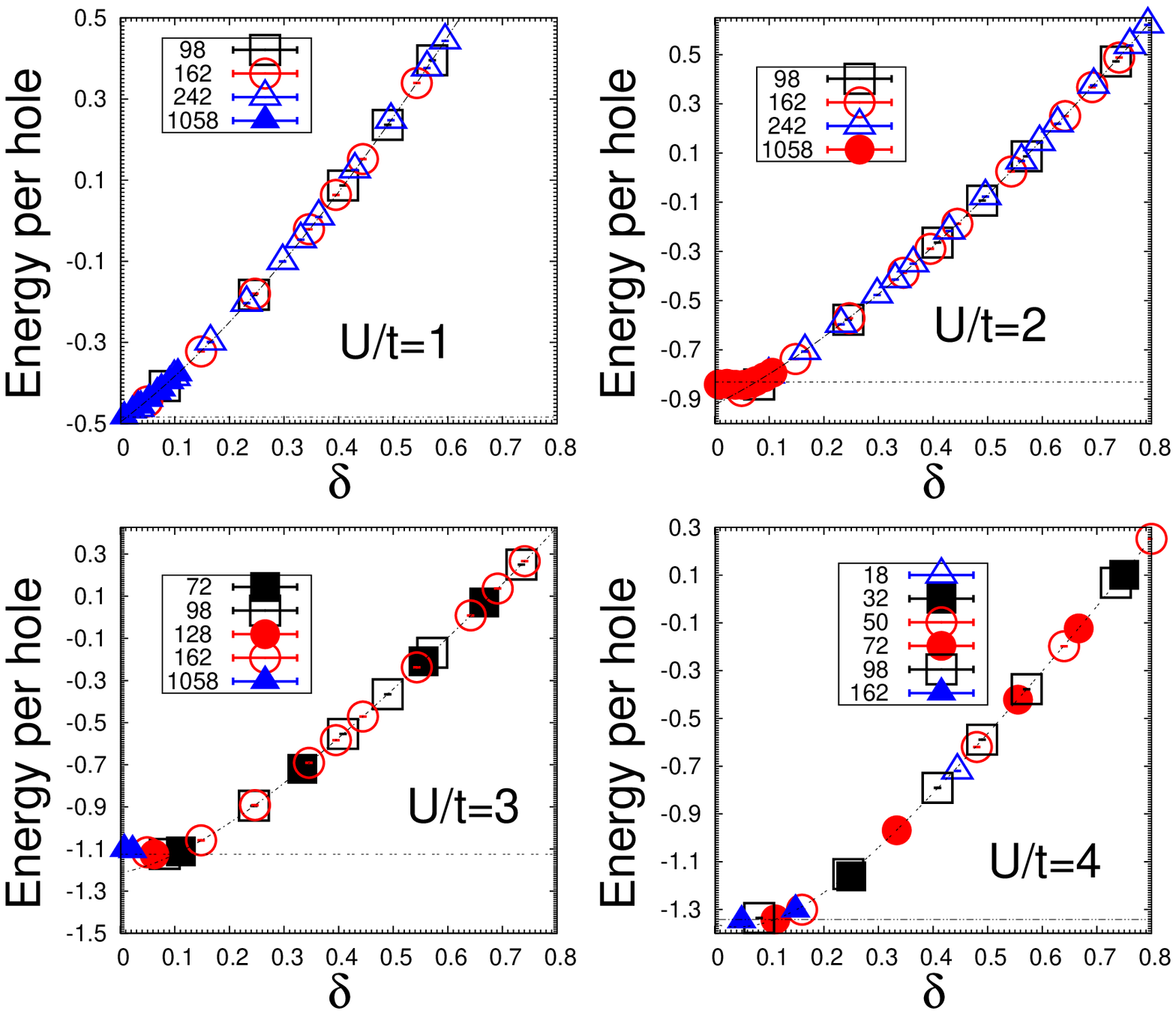}
\caption{ \label{spectrum} Energy per hole for various $U/t$ and cluster sizes. The curved lines 
are fit to the data with $5^{th}$ order polynomials (see main text).
The horizontal lines are the lowest hole energies estimated at the smallest available
$\delta$ on the largest clusters.
}\label{allu}
\end{figure}

The results for the hole energy for $U/t=1,2,3,4$ are summarized in Fig.(\ref{allu}).
The Trotter discretization time is chosen to have a positive systematic error (the values reported are variational upper bound of the energy) in the energy per site 
within $2 \times 10^{-4} t$, namely 
$\Delta t \times t={1\over 4},{1\over 6},{1\over 8},{1\over 10}$ for 
$U/t=1,2,3,4$, respectively. This error, being systematic and similar for nearby  dopings, 
 is expected to cancel out for the computation of the hole 
energy at small dopings, that, in this case, is affected only by the statistical error. 
Notice that, thanks to the symmetrized expression of the short time propagator:
\begin{equation}
\exp(-\Delta \tau H) \simeq \exp(-{\Delta \tau \over 2 }  K ) \exp ( -\Delta \tau V ) 
\exp(-{\Delta \tau \over 2 }  K )  
\end{equation}
the Trotter error in the energy becomes almost negligible if measured
after long imaginary time projection, as it is scaling as $(\Delta \tau)^4$,\footnote{
At leading order the approximate ground state wave function $|\psi^{GS}_{\Delta \tau}\rangle$  as a function of 
$\Delta \tau$ can be written as $|\psi^{GS}_{\Delta \tau}\rangle= 
|\psi^{GS}\rangle + \Delta \tau^2 |\psi^\prime \rangle + O (\Delta \tau^3)$ 
where $|\psi^\prime \rangle$ is orthogonal to the exact ground state 
$|\psi^{GS}\rangle$. Thus it is simple to show that 
the approximate ground state energy $E(\Delta \tau)= 
{ \langle \psi^{GS}_{\Delta \tau} | H | \psi^{GS}_{\Delta \tau} \rangle \over 
\langle \psi^{GS}_{\Delta \tau} | \psi^{GS}_{\Delta \tau} \langle }$ does not 
contain the leading $O(\Delta \tau^2)$ order in the expansion, and that, 
with simple inspection, the 
first non zero contribution is vanishing as $\Delta \tau^4$}   
 and is also variational in the "Var" case mentioned in the paper 
because it corresponds to an energy expectation value of a given state.

\begin{table}[h]
\begin{tabular}{|l|l|l|l|l|l|}
\hline
 L & $N_h$ &  $U/t=1$ & $U/t=2$ & $U/t=3$ & $U/t=4$  \\
\hline
18  & 0 & -1.383984(25)          & -1.174357(42)         & -.995837(80)          & -.851727(22) \\
18  & 8 & -1.316280(18)          & -1.258224(50)         & -1.210676(57)         & -1.171736(17) \\
18  & 16 & -.404113(4)   & -.402104(12)          & -.400564(15)          & -.399349(4) \\
\hline
32  & 0 & -1.383759(19)         & -1.173957(49)         & -.996540(66)  & -.851421(35) \\
32  & 8 & -1.427314(18)         & -1.315238(40)         & -1.220670(61)         & -1.141815(28) \\
32  & 24 & -.802289(8)  & -.791310(14)  & -.782523(24)  & -.775319(17) \\
\hline
50  & 0 & -1.383931(9)  & -1.175175(26)           & -.999904(34)       & -.856989(29) \\
50  & 8 & -1.432718(9)  & -1.289485(22)          & -1.167422(25)        & -1.065190(50) \\
50  & 24 & -1.279282(8)         & -1.228975(19)          & -1.187942(20)        & -1.154477(15) \\
50  & 40 & -.671976(3)  & -.665129(9)    & -.659723(10)         & -.655354(6) \\
50  & 48 & -.154689(1)  & -.154454(2)    & -.154276(2)  & -.154139(2) \\
\hline
72  & 0 & - & - & -1.000386(41)         & -.857926(40) \\
72  & 8 & - & - & -1.123689(54)         & -1.007007(30) \\
72  & 24 & - & - & -1.239880(23)        & -1.180751(37) \\
72  & 40 & - & - & -1.116184(17)        & -1.092080(30) \\
72  & 48 & - & - & -.953056(13)         & -.939974(24) \\
72  & 64 & - & - & -.400496(6)  & -.399212(10) \\
\hline
98  & 0 & -1.383884(17)          & -1.175669(28)         & -1.001029(53)         & -.859353(19) \\
98  & 8 & -1.417291(113)         & -1.242952(157)        & -1.093107(165)        & -.968388(37) \\
98  & 24 & -1.428730(86)         & -1.315649(123)        & -1.220505(82)         & -1.141423(43) \\
98  & 40 & -1.348419(86)         & -1.281963(95)         & -1.227174(45)         & -1.182168(17) \\
98  & 48 & -1.268210(58)         & -1.219844(107)        & -1.180124(43)         & -1.147767(19) \\
98  & 56 & -1.158019(59)         & -1.124638(87)         & -1.097401(41)         & -1.075118(17) \\
98  & 72 & -.839414(50)          & -.826997(55)          & -.817283(28)          & -.809193(10) \\
98  & 80 & -.626355(26)          & -.620616(36)          & -.616013(20)          & -.612351(8) \\
98  & 88 & -.374069(17)          & -.372394(30)          & -.371059(12)          & -.370032(4) \\
98  & 96 & -.080244(4)   & -.080195(5)   & -.080132(7)   & -.080112(1) \\
\hline
128  & 0 & - & - & -1.001331(40) & - \\
128  & 8 & - & - & -1.071665(50) & - \\
\hline
162  & 0 & -1.383989(68)         & -1.175761(25)        & -1.001315(46)          & -.859880(39) \\
162  & 8 & -1.405966(77)         & -1.217640(156)       & -1.056538(36)          & -.926461(650) \\
162  & 24 & -1.431606(55)        & -1.284265(104)       & -1.158330(69)          & -1.052891(110) \\
162  & 40 & -1.428448(56)        & -1.316032(108)       & -1.221628(340)  & - \\
162  & 56 & -1.391377(56)        & -1.308639(113)       & -1.240138(69) & - \\
162  & 64 & -1.358795(55)        & -1.289391(99)        & -1.231708(36) & - \\
162  & 72 & -1.316351(55)        & -1.258192(80)        & -1.210626(37) & - \\
162  & 88 & -1.199681(41)        & -1.161410(84)        & -1.130263(28) & - \\
162  & 104 & -1.037246(42)       & -1.014225(82)        & -.995547(21) & - \\
162  & 112 & -.937604(37)        & -.920560(60)         & -.907076(19)  & - \\
162  & 120 & -.824934(29)        & -.813331(64)         & -.803772(14)  & - \\
162  & 136 & -.558768(20)        & -.554421(39)         & -.551001(11)  & - \\
162  & 144 & -.404113(14)        & -.402095(22)         & -.400539(9)  & - \\
162  & 152 & -.234318(10)        & -.233719(20)         & -.233277(7)  & - \\
162  & 160 & -.048875(2)         & -.048852(4)  & -.048838(1)  & - \\
\hline
242  & 0 & -1.383905(64)         & -1.175775(63) & -1.001435(26) & -.860127(38) \\
242  & 8 & -1.399151(64)         & -1.203729(159)  & - & - \\
242  & 24 & -1.422228(55)        & -1.255013(59)  & - & - \\
242  & 40 & -1.433352(67)        & -1.292197(45)  & - & - \\
242  & 56 & -1.430992(59)        & -1.313451(30)  & - & - \\
242  & 72 & -1.413774(59)        & -1.317353(24)  & - & - \\
242  & 80 & -1.399206(48)        & -1.312518(24)  & - & - \\
242  & 88 & -1.380472(62)        & -1.302740(23)  & - & - \\
242  & 104 & -1.330079(51)       & -1.268556(21)  & - & - \\
242  & 120 & -1.261094(49)       & -1.213780(19)  & - & - \\
242  & 136 & -1.172439(51)       & -1.137193(20)  & - & - \\
242  & 144 & -1.120353(43)       & -1.090458(17)  & - & - \\
242  & 152 & -1.062959(31)       & -1.037906(15)  & - & - \\
242  & 168 & -.931407(31)        & -.914761(16)  & - & - \\
242  & 184 & -.776579(27)        & -.766544(11)  & - & - \\
242  & 192 & -.690070(20)        & -.682683(10)  & - & - \\
242  & 200 & -.597216(21)        & -.592090(8)  & - & - \\
242  & 216 & -.392008(14)        & -.390109(5)  & - & - \\
242  & 224 & -.279298(10)        & -.278419(4)  & - & - \\
242  & 232 & -.159624(6)         & -.159371(2)  & - & - \\
242  & 240 & -.032829(2)         & -.032820(0)  & - & - \\
\hline
1058  & 0 & -1.383674(124)       & -1.175469(92)         & -1.001412(52)  & -  \\
1058  & 8 & -1.387331(148)       & -1.181826(91)         & -1.009482(123)  & -  \\
1058  & 24 & -1.394317(55)       & -1.194489(118)        & -1.026282(233)  & -  \\
1058  & 40 & -1.400945(57)       & -1.207317(101)        &  - & - \\
1058  & 56 & -1.406985(56)       & -1.220074(63)         & - & - \\
1058  & 72 & -1.412511(61)       & -1.232271(121)        & - & - \\
1058  & 80 & -1.415030(65)       & -1.238027(63)         & - & - \\
1058  & 96 & -1.419688(53)       & -1.249151(56)         & - & - \\
1058  & 112 & -1.423659(69)      & -1.259463(47)         & - & - \\
\hline
\end{tabular}
\caption{ Energy per site for several $U/t$ and number of holes by using MBPC. 
For the largest cluster $L=1058$ or $U/t=3,4$ 
there may be a systematic bias in imaginary time convergence of a few standard deviations. This should be compensated in the energy per hole calculation at small doping, because they involve energy differences (converging faster) computed  with {\em the same} imaginary 
time projection. } 
\end{table}

For the smallest doping and largest clusters and  $U/t>2$ values, 
in order to have converged energies, it is necessary to 
optimize the trial function.
Only at $U/t=4$ we have found convenient to use a mean field Hamiltonian with a small 
 $d-$wave superconducting order parameter.
In general $\Delta_{BCS}^{x^2-y^2}$ and especially 
$\Delta_{AF}$ are useful in the small doping region.